\documentclass[lettersize,journal]{IEEEtran}
\usepackage{amsmath,amsfonts}
\usepackage{algorithmic}
\usepackage{algorithm}
\usepackage{array}
\usepackage[caption=false,font=normalsize,labelfont=sf,textfont=sf]{subfig}
\usepackage{textcomp}
\usepackage{stfloats}
\usepackage{url}
\usepackage{verbatim}
\usepackage{graphicx}
\usepackage{cite}
\begin{document}
\title{Large Language Models Meet Bayesian Optimization: Towards Self-evolving Wireless Digital Twins}
\author{Yingzhuo Sun, ~\IEEEmembership{Student Member,~IEEE,}
Haoyun Li, ~\IEEEmembership{Student Member,~IEEE,}
Ming Xiao, ~\IEEEmembership{Senior Member,~IEEE,} 
Francisco J. Ferrão, and
Berna \"{O}zbek, ~\IEEEmembership{Senior Member,~IEEE} 
\thanks{Y. Sun, H. Li, and M. Xiao are with the Division of Information Science and Engineering, KTH Royal Institute of Technology, 100 44 Stockholm, Sweden
(e-mail: \{yingzhuo, haoyunl, mingx\}@kth.se).}
\thanks{F. J. Ferrão is with Data Machine Elite, 2580-654 Lisbon, Portugal (e-mail: francisco.ferrao@datamachineelite.com).}
\thanks{B. \"{O}zbek with the Department of Electrical and 
Electronics Engineering, \.{I}zmir Institute of Technology, 35430
\.{I}zmir, T\"{u}rkiye (e-mail: bernaozbek@iyte.edu.tr).}
}
\maketitle
\begin{abstract}
Wireless digital twins (DTs) are increasingly integrated with artificial intelligence to support online modeling, prediction, and network optimization. However, most existing approaches perform adaptation within predefined learning or optimization structures, which may become unsuitable when wireless tasks and operating conditions change. In this article, we investigate the integration of large language models (LLMs) and Bayesian optimization (BO) to extend wireless DTs from online adaptation towards algorithm-level self-evolution. We introduce an LLM-assisted BO framework in which BO provides a stable and sample-efficient optimization backbone, while the LLM determines whether selected BO components should be retained or adapted based on the current task and accumulated optimization experience. In simulations, we apply the framework to Gaussian process (GP) kernel discovery for a multi-user wireless beam-selection problem. Simulation results show that the LLM-discovered kernel improves the final BO performance by 19.7\% on the training tasks compared with the strongest standard kernel baseline. The results demonstrate the potential of combining LLM flexibility with structured BO for self-evolving wireless DTs.
\end{abstract}
\begin{IEEEkeywords}
Wireless digital twin, large language models, Bayesian optimization, algorithm adaptation.
\end{IEEEkeywords}
\section{Introduction}
Digital twins (DTs) have emerged as a promising approach for modeling, monitoring, and controlling complex wireless networks. By maintaining a digital representation of the physical system, a wireless DT can support network-state estimation, performance prediction, resource optimization, and virtual testing of candidate decisions. As wireless systems grow more complex, these capabilities are becoming increasingly important for network management and autonomous operation. However, conventional model-driven DTs usually rely on predefined models. The accuracy and adaptability may degrade when the wireless environment becomes highly dynamic or difficult to model. To improve modeling and adaptation, recent studies have introduced data-driven methods such as deep learning and reinforcement learning. These artificial intelligence (AI)-driven DTs can support online model updating, adaptive control, and closed-loop interaction between the digital and physical domains \cite{DigitalTwin1}.
Despite these advances, an important limitation remains. Most existing AI-driven wireless DTs perform adaptation within predefined learning and optimization structures. Model parameters, network configurations, or control policies can be updated online, while the underlying learning model, optimization strategy, or decision mechanism is usually designed in advance. This approach works well when operating conditions remain close to those considered during algorithm design or training. However, future wireless networks may involve heterogeneous tasks, changing objectives, and varying network scales. In such cases, a fixed optimization strategy may no longer be effective. Therefore, the next step is not only to adapt the network state or policy, but also to adapt how the optimization itself is performed.
Bayesian optimization (BO) is well suited to this DT setting. BO uses observations from the physical system to update its surrogate model and generate optimization decisions. This iterative process naturally supports the closed-loop operation of wireless DTs. BO has therefore been applied to several wireless problems, including radio resource management, power control, beam alignment, and meta-optimization \cite{BOWireless1,BOWireless2,BOWireless3,BOWireless4}. However, the BO framework itself is still usually predefined, including initialization, surrogate model, kernel, and acquisition function. When the wireless task or operating scenario changes, the same BO configuration may no longer be suitable.
Large language models (LLMs) offer a complementary way to address this limitation. LLMs can understand task descriptions, reason over previous experience, and generate new algorithmic structures or executable code. These capabilities make them suitable for adapting the BO framework to new wireless tasks. Recent studies have already shown the use of LLMs for wireless-network reasoning, configuration, optimization, and agent-based control \cite{LLMWireless1,LLMWireless2,LLMWireless3}. LLMs have also been used to generate and refine heuristics and algorithmic structures \cite{LLMAlgorithm1,LLMAlgorithm2}, and more recent studies have introduced LLMs into BO for reasoning-guided optimization and BO-component generation \cite{LLMBO1,LLMBO2,LLMBO3,LLMBO4}. These developments suggest that LLMs can complement BO by adapting its components when wireless tasks and environments change, thereby extending the DT from online model and policy adaptation toward algorithm-level self-evolution.
Motivated by these observations, we investigate the integration of LLMs and BO for wireless DTs. The main idea is to combine the sample-efficient black-box optimization of BO with the reasoning and algorithm-generation capabilities of LLM. Instead of using a fixed BO pipeline for all wireless tasks, the proposed framework allows the system to collect previous task, algorithm, and performance information, and use this experience to determine whether the current BO configuration should be retained or adapted. When adaptation is required, the LLM can generate new BO components and return them to the optimization loop. In this way, the DT can extend conventional online adaptation from the state or policy level to the optimization-mechanism level.
The main contributions are summarized as follows.
\begin{itemize}
    \item We discuss the complementary roles of LLMs, BO, and wireless intelligence, and introduce the concept of algorithm-level self-evolution for wireless DTs. Different from conventional online adaptation, the focus is on adapting the optimization mechanism itself when wireless tasks and operating conditions change.
    \item We propose an LLM-assisted BO framework for self-evolving wireless DTs. The framework integrates wireless-task interaction, BO execution, experience storage, prompt construction, LLM-based adaptation decisions, and BO-component generation into a closed loop.
    \item We provide a proof-of-concept study on LLM-assisted BO component discovery. Kernel discovery is considered a representative example to evaluate whether LLM-generated BO components can improve optimization performance across wireless tasks.
\end{itemize}
The remainder of this article is organized as follows. Section II discusses the motivation for combining LLMs and BO in wireless DTs. Section III introduces the proposed framework. Section IV presents the simulation study. Section V discusses open challenges. Section VI concludes the article.
\section{From Online Adaptation to Self-Evolving Wireless Digital Twins}
\begin{figure}[tb]
	\centerline{\includegraphics[width=0.5\textwidth]{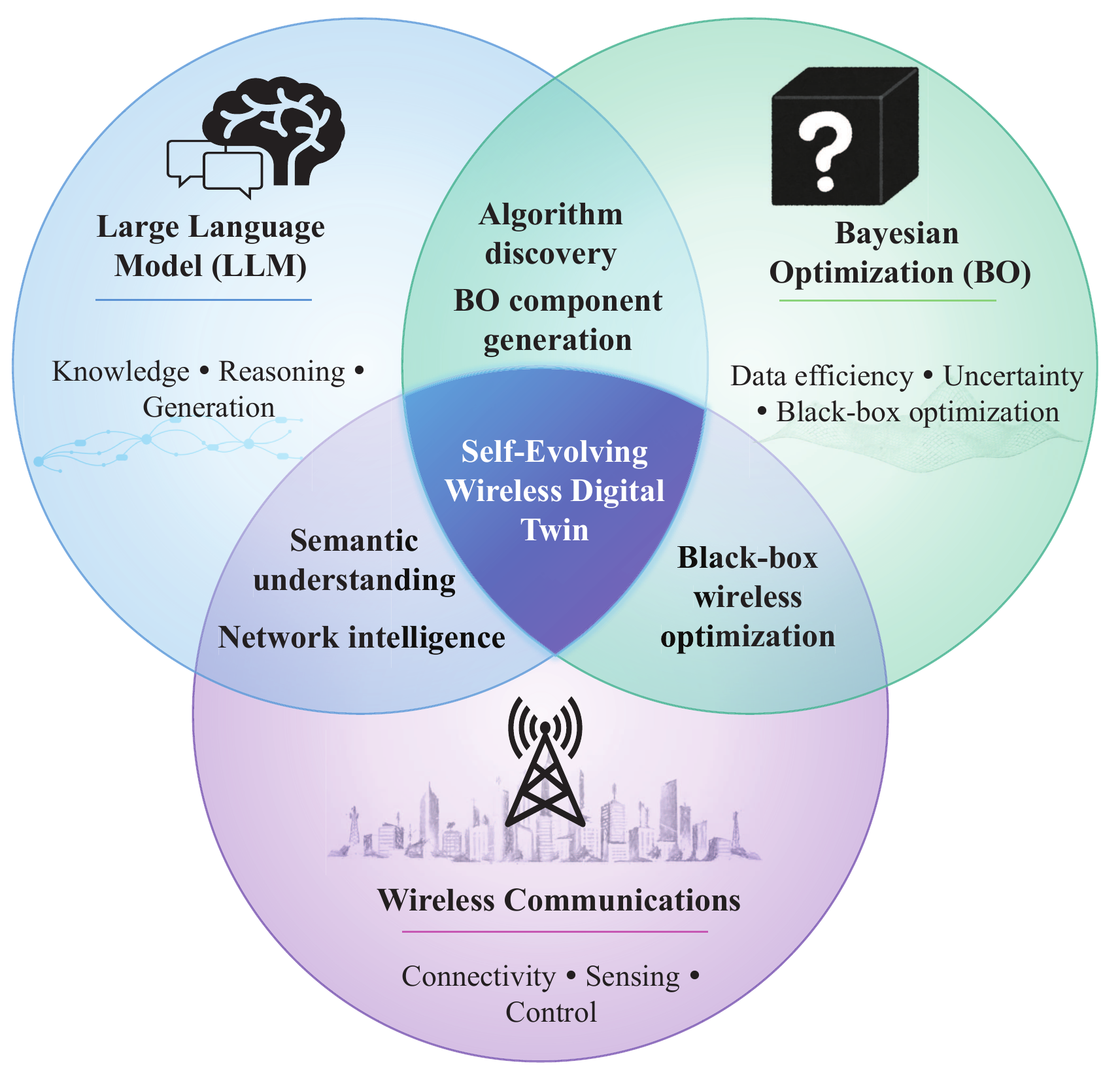}}
	\caption{Connections of LLMs, Bayesian optimization, and wireless systems towards self-evolving wireless digital twins.}
	\label{fig:Literature_map}
\end{figure}
LLMs and BO provide complementary capabilities for intelligent wireless optimization. LLMs can offer strong task-understanding and algorithm-generation capabilities, while BO provides a structured and sample-efficient approach to black-box optimization. Their different strengths and limitations motivate their integration for wireless DTs.
\subsection{Large Language Models for Wireless Intelligence}
LLMs are particularly useful when tasks cannot be fully specified by a fixed input-output mapping. They can interpret task descriptions, reason over contextual information, and generate task-specific decisions, algorithms, or executable code. These capabilities matter for wireless systems, where objectives, constraints, network scales, and operating conditions can vary across tasks. Recent studies have therefore explored LLMs for wireless reasoning, configuration, optimization, and agent-based control \cite{LLMWireless1,LLMWireless2,LLMWireless3}. Large sequence models have also been investigated for wireless signal processing, using in-context learning to perform MIMO equalization across different user distributions\cite{OJCom}. LLMs have also demonstrated the ability to generate and refine heuristics and algorithmic structures \cite{LLMAlgorithm1,LLMAlgorithm2}.
However, directly using an LLM to generate optimization algorithms also introduces a concern. Although the generated algorithms can have explicit structures and executable implementations, the search space is highly flexible, and the LLM may substantially change the overall optimization procedure across tasks. As a result, different tasks may be solved by very different algorithmic structures, making the optimization process more difficult to constrain, compare, and reuse. For wireless DTs, where the optimization mechanism is expected to operate repeatedly across changing tasks, it is desirable to retain a stable optimization framework while allowing only selected components to evolve.
\subsection{Bayesian Optimization for Wireless Black-Box Problems}
BO provides such a structured framework for problems in which the objective function is difficult to model and expensive to evaluate. This setting is common in wireless systems because system performance can depend on complex interference, channel conditions, traffic, mobility, and coupled resource decisions. BO can use a limited number of system observations to update a surrogate model and sequentially select new decisions, making it suitable for sample-efficient interaction with wireless systems \cite{BOWireless1,BOWireless2,BOWireless3,BOWireless4}.
More importantly, BO has an explicit modular structure, which makes the optimization process interpretable and allows specific behaviors to be traced to individual algorithmic choices. However, these choices are commonly specified in advance. A BO configuration developed for one task may therefore become unsuitable when the objective, decision space, or operating scenario changes. BO can adapt its model to new observations, but it does not naturally adapt its own optimization structure.
\subsection{Combining LLMs and Bayesian Optimization}
LLMs and BO offer complementary capabilities. LLMs can generate task-specific algorithms, while BO provides a stable optimization workflow with predefined components. Their combination allows selected BO components to evolve without redesigning the overall optimization procedure. For example, the LLM can generate or modify the initialization method, kernel function, AF, or candidate-search strategy. Each component can have a predefined interface to ensure compatibility with the BO engine. Recent studies on LLM-assisted BO and BO-component generation have demonstrated the potential of such structured interaction \cite{LLMBO1,LLMBO2,LLMBO3,LLMBO4}.
\subsection{Towards Self-Evolving Wireless Digital Twins}
This structured combination provides a natural path toward self-evolving wireless DTs. Conventional online adaptation changes system states, model parameters, or control decisions within a predefined algorithm. In contrast, algorithm-level self-evolution allows selected optimization components to change according to the current task and accumulated experience while retaining the overall BO workflow. Fig.~\ref{fig:Literature_map} summarizes the complementary roles of wireless DTs, BO, and LLMs in supporting this process.
\section{An LLM-Assisted BO Framework for Wireless Digital Twins}
\begin{figure*}[tb]
	\centerline{\includegraphics[width=1\textwidth]{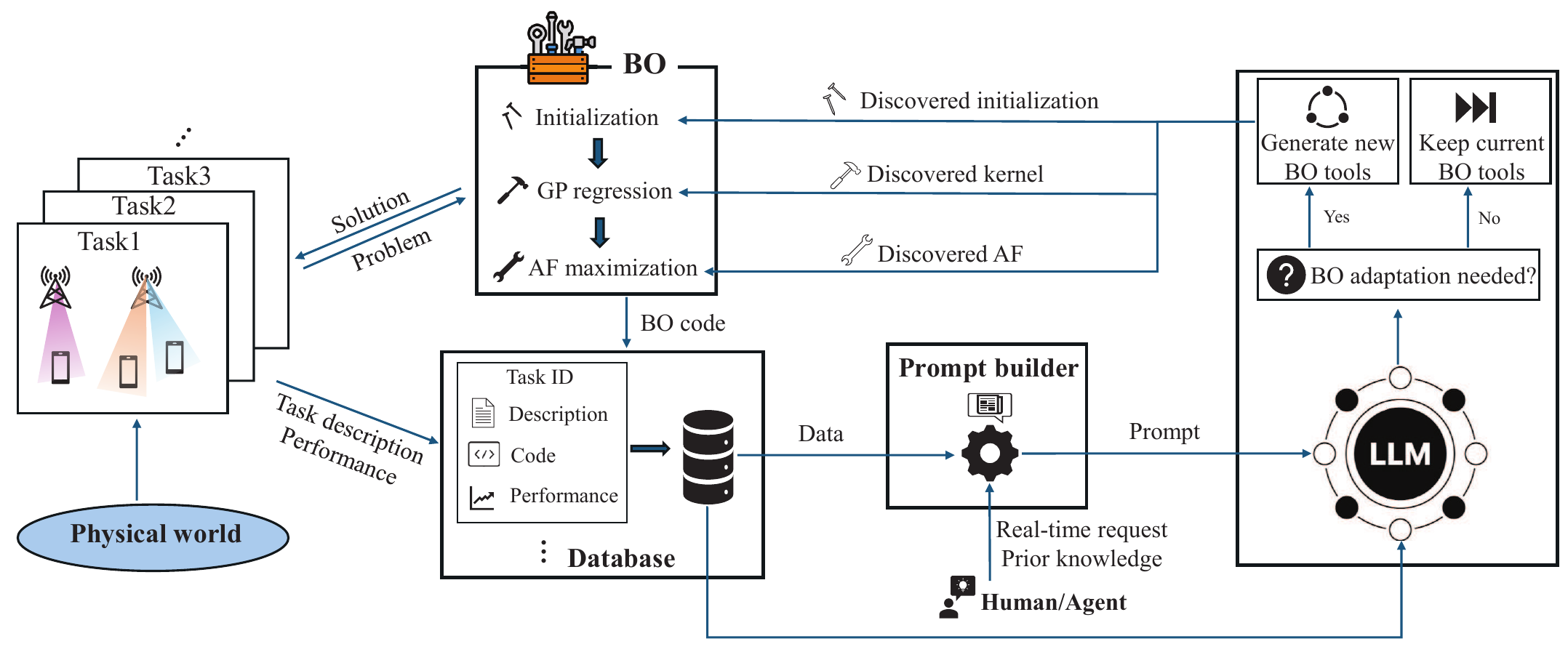}}
	\caption{The proposed framework for self-evolving wireless digital twins.}
	\label{fig:DT_framework}
\end{figure*}
Fig.~\ref{fig:DT_framework} illustrates the proposed framework for self-evolving wireless DTs. It consists of four main modules: a BO engine, a database, a prompt builder, and an LLM-driven adaptation module. These modules form a closed loop that connects wireless optimization, experience accumulation, and BO-component adaptation.
\subsection{Bayesian Optimization Engine}
The BO engine is responsible for solving the wireless optimization task and interacting with the physical system. As shown in Fig.~\ref{fig:DT_framework}, different tasks may arrive over time. They may correspond to changes in the optimization objective, network size, user distribution, or channel conditions. For each task, the wireless system provides the optimization problem to the BO engine, while BO iteratively returns candidate solutions for evaluation.
The BO procedure is divided into three main stages: initialization, Gaussian process (GP) regression, and acquisition-function (AF) maximization. These stages determine how BO starts, models the objective, and selects new candidates, respectively. Initialization determines the observations available at the beginning of BO. GP regression constructs and updates the surrogate model using the evaluated samples. The kernel function determines the covariance structure of the GP surrogate. Based on the posterior information provided by the GP, the AF balances the predicted performance and uncertainty to select the next candidate for evaluation. The newly observed performance is then used to update the GP surrogate model.
This iterative process supports physical–digital interaction without requiring a fully accurate analytical model of the wireless environment.
More importantly, the BO engine is modular. The overall sequence of initialization, GP surrogate model fitting, and AF maximization can remain unchanged, while individual components can be replaced when the current design becomes unsuitable. In Fig.~\ref{fig:DT_framework}, the LLM may return a discovered initialization method, GP kernel, or AF to the BO engine. This modularity provides the interface required for controlled algorithm-level evolution without redesigning the complete optimization workflow.
\subsection{Experience Database}
The experience database stores records from previous wireless optimization tasks. As illustrated in Fig.~\ref{fig:DT_framework}, each record contains the task identity, task description, BO code, and observed optimization performance. These records establish the relationship between wireless tasks, algorithm implementations, and their performance.
When a new task arrives, relevant records can be retrieved based on task similarity, algorithm structure, or observed performance. The database is continuously updated after new BO components are evaluated, providing accumulated experience for subsequent LLM-based adaptation.
\begin{figure*}[tb]
	\centerline{\includegraphics[width=0.9\textwidth]{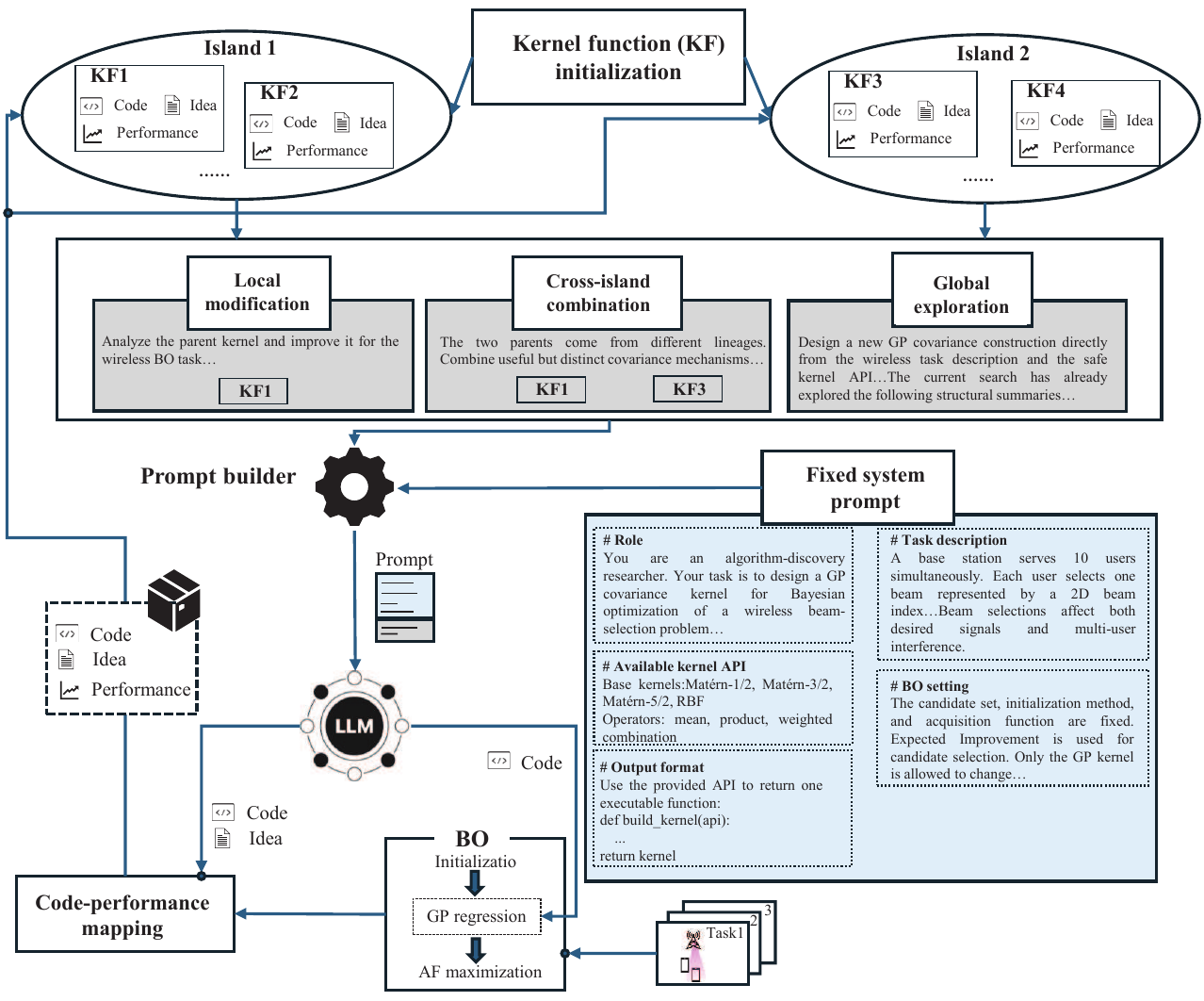}}
	\caption{LLM-assisted GP kernel discovery with island-based evolutionary search and BO performance feedback.}
	\label{fig:Algorithm}
\end{figure*}
\subsection{Prompt Builder}
The prompt builder combines the current wireless-task description with relevant records retrieved from the experience database. Prior knowledge or additional instructions from a human or another agent can also be included. The resulting prompt provides the task information and optimization experience required for LLM-based adaptation.
\subsection{LLM-Driven BO Adaptation}
The LLM-driven module is responsible for algorithm-level adaptation of the BO engine. After receiving the structured prompt, the LLM evaluates whether the current BO configuration remains suitable for the new wireless task. When additional historical information is needed, it can also access the experience database to retrieve relevant task--algorithm--performance records.
If the current BO configuration remains suitable, the existing BO tools are retained. Otherwise, the LLM identifies which component should be modified and enters the BO-tool generation stage. In the framework shown in Fig.~\ref{fig:DT_framework}, the considered components include the initialization method, GP kernel, and acquisition function. The LLM can therefore generate a new component according to the current task and the accumulated optimization experience.
The generated component is returned to the corresponding stage of the BO engine through a predefined interface. In this way, the LLM does not redesign the complete optimization procedure, but adapts selected BO components while the overall BO workflow remains unchanged.
Therefore, the proposed framework separates two levels of adaptation. BO performs online adaptation within a given optimization configuration through continuous surrogate updating and candidate selection. The LLM operates at a slower algorithmic level and modifies selected BO components when the existing configuration is no longer suitable. By combining these two loops, the wireless DT can retain a stable BO backbone while allowing its optimization mechanism to evolve across changing tasks.
\section{Simulation Experiments}
To demonstrate the feasibility of the proposed framework, we consider GP kernel discovery as a representative example of LLM-assisted BO component adaptation. The purpose of the simulation is not to implement every module in the proposed framework. Instead, we keep the overall BO procedure fixed and make the LLM evolve only the kernel function. 
\begin{figure}[tb]
    \centerline{\includegraphics[width=0.5\textwidth]{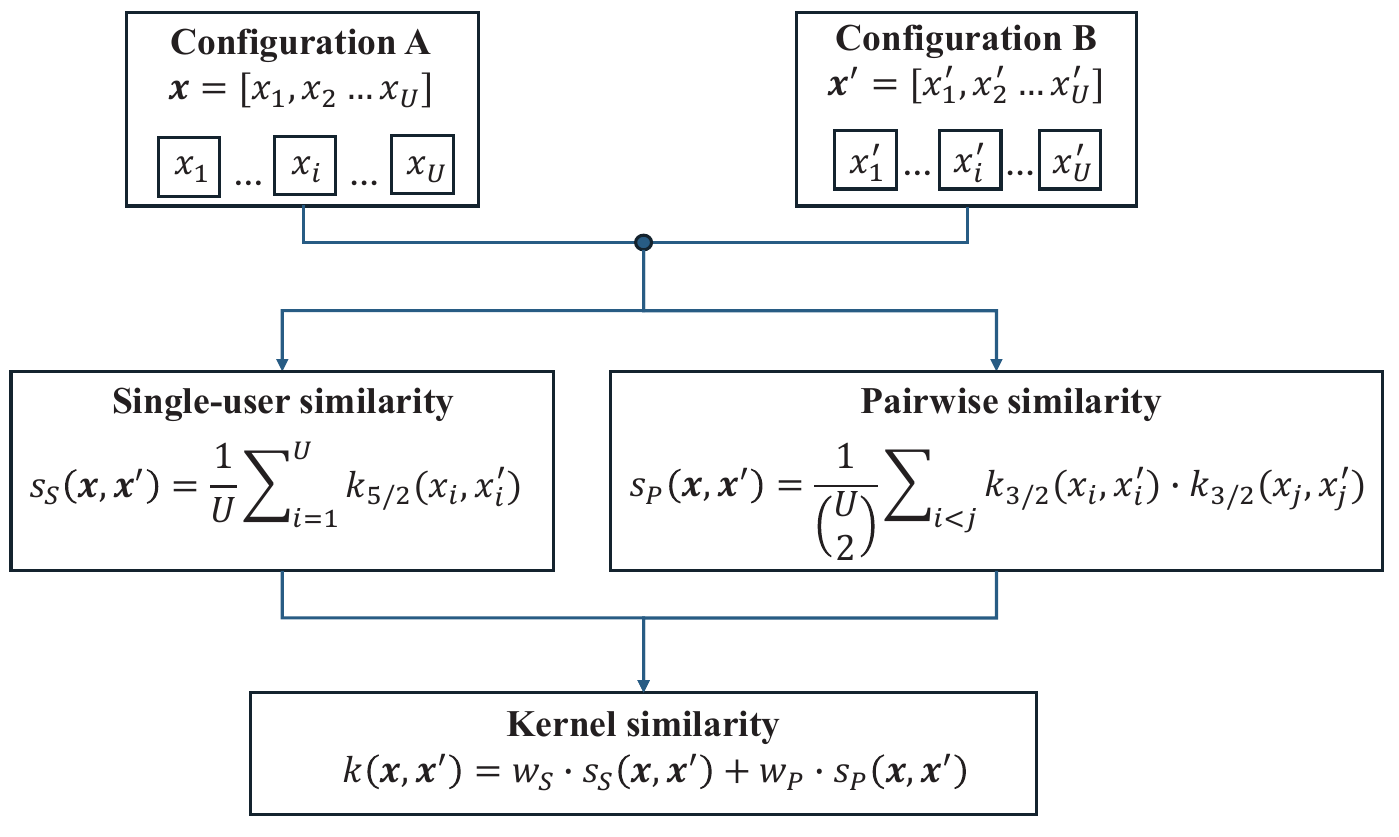}}
    \caption{Structure of the GP kernel discovered by the LLM for the multi-user beam-selection task.}
    \label{fig:discovered_kernel}
\end{figure}
\begin{figure}[tb]
    \centerline{\includegraphics[width=0.5\textwidth]{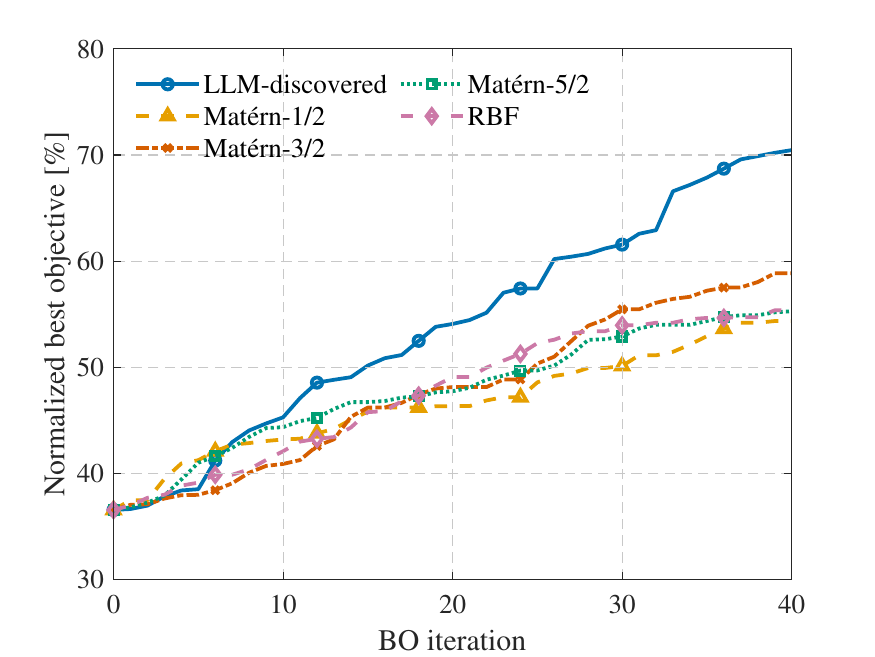}}
    \caption{BO convergence with the LLM-discovered and standard GP kernels on the training tasks.}
    \label{fig:kernel_performance}
\end{figure}
\subsection{Wireless Optimization Task}
We consider a multi-user downlink beam-selection problem. The wireless channel data are generated using Sionna under an urban scenario. A base station equipped with an $8\times8$ uniform planar array simultaneously serves 10 users at a carrier frequency of 3.5 GHz. A two-dimensional discrete Fourier transform (DFT) codebook with 64 candidate beams is used, and each user is assigned one beam. The total transmit power is equally divided among the users.
Each beam-selection configuration contains the beam choices of all 10 users. Since each beam is represented by a two-dimensional beam index, a configuration is mapped to a 20-dimensional input for the GP surrogate model. The optimization objective is the sum spectral efficiency, given by $f(\mathbf{x})=\sum_{i=1}^{U}\log_2(1+\mathrm{SINR}_i(\mathbf{x}))$
where $U=10$ is the number of users, $\mathbf{x}$ represents their beam-selection configuration, and $\mathrm{SINR}_i(\mathbf{x})$ is the signal-to-interference-plus-noise ratio of user $i$. The SINR depends on the desired beam gain, interference from beams assigned to other users, and receiver noise. These coupled effects make the problem suitable for evaluating different GP covariance structures.
A total of 50 wireless tasks are generated. Each task corresponds to a different realization of the users and channels. Among them, 30 tasks are used for LLM-based kernel discovery, while the remaining 20 tasks are reserved for testing. For each task, all kernels use the same 10 randomly selected initial observations, followed by 40 BO iterations. The initialization procedure, GP fitting procedure, and acquisition function are kept identical for all kernels.
\subsection{LLM-Based Kernel Discovery}
For the experiments, the general adaptation framework is simplified to kernel discovery. The BO structure, initialization, and AF are fixed, and expected improvement (EI) is used as the AF throughout the experiment. Therefore, differences in BO performance originate from the GP kernel rather than from other BO components.
The kernel discovery design follows the evolutionary search framework in \cite{LLMAlgorithm1,LLMAlgorithm2}. Fig.~\ref{fig:Algorithm} illustrates the LLM-assisted GP kernel discovery process used in the simulation. Candidate kernels are maintained in two islands together with their code, design idea, and observed BO performance. New kernels are generated through local modification, cross-island combination, or global exploration. The corresponding discovery instruction is combined with a fixed system prompt and provided to the LLM, which returns a new kernel program. The generated kernel is then evaluated by the BO engine, and its performance is fed back to update the kernel population.
Fig.~\ref{fig:discovered_kernel} illustrates the structure of the discovered kernel. For two beam configurations, the kernel contains two similarity components. The first applies a shared Mat\'ern-5/2 kernel to the 2D beam index of each user and then averages the resulting similarities across users. The second applies a shared Mat\'ern-3/2 kernel to the 2D beam indices of two users and combines the two similarities multiplicatively, capturing pairwise interactions. The two components are finally combined using learned positive weights. This structure is consistent with the wireless task: the first component captures similarity in individual beam selections, while the second reflects coupling between user pairs caused by interference. Here, $w_s$ and $w_p$ are the learned positive weights of the single-user and pairwise similarity components, respectively, with $w_s+w_p=1$.
\subsection{Performance Evaluation}
We compare the discovered kernel with four standard GP kernels: Mat\'ern-1/2, Mat\'ern-3/2, Mat\'ern-5/2, and radial basis function (RBF). All methods use the same initialization, AF, and evaluation budget.
Fig.~\ref{fig:kernel_performance} shows the average BO convergence over the 30 training tasks. All kernels start from the same initial performance because they use identical initial observations. As additional BO evaluations are performed, the LLM-discovered kernel outperforms the standard kernels gradually. After 40 BO iterations, it reaches a normalized best objective of 70.46\%, whereas the strongest standard kernel, Mat\'ern-3/2, reaches 58.85\%. This corresponds to an improvement of 19.7\% relative to the strongest baseline. More importantly, the advantage is retained on the 20 test tasks, which are not used during kernel discovery. As summarized in Table~\ref{tab:kernel_performance}, the discovered kernel achieves a normalized final objective of 0.6482, compared with 0.5962 for the strongest standard baseline. The corresponding improvement is 8.7\%. The improvement on unseen wireless realizations indicates that the discovered covariance structure is not limited to the tasks used for kernel discovery.
These results provide a proof of concept for the main idea of the proposed framework, in which the LLM searches for a task-relevant component within a fixed BO structure. The discovered kernel also outperforms all four standard kernels on test tasks. This shows the potential of using accumulated task--algorithm--performance experience to evolve BO components for changing wireless optimization problems.
\begin{table}[t]
\centering
\caption{Comparison of the discovered kernel with standard kernels.}
\label{tab:kernel_performance}
\renewcommand{\arraystretch}{1.15}
\begin{tabular}{lcc}
\hline
GP Kernel & Training Tasks & Test Tasks \\
\hline
{\bfseries LLM-discovered} 
    & {\bfseries 0.7046} 
    & {\bfseries 0.6482} \\
Mat\'ern-1/2 & 0.5434 & 0.5717 \\
Mat\'ern-3/2 & 0.5885 & 0.5962 \\
Mat\'ern-5/2 & 0.5526 & 0.5134 \\
RBF          & 0.5535 & 0.5217 \\
\hline
Improvement 
    & 19.7\% 
    & 8.7\% \\
\hline
\end{tabular}
\end{table}
\section{Open Challenges and Research Directions}
Although the proposed framework provides a possible path toward algorithm-level self-evolution, several challenges and potential research directions remain before it can be applied to practical wireless DTs in long-term operation.
\subsection{Adaptation under Task Changes}
The current simulation considers multiple wireless tasks generated from the same general system model. Although the discovered kernel improves performance on test tasks, these tasks still follow a similar underlying distribution. In practical wireless DTs, task changes may arise from changes in the wireless system, such as channels, traffic, mobility, or network scale, as well as changes in the optimization objective or constraints.
A key research challenge is therefore to determine whether the current BO configuration remains suitable after such changes. When the task changes only moderately, previously discovered components may still be reusable. Larger changes may require adaptation of the initialization method, surrogate model, kernel, or AF. Future work should develop systematic criteria for deciding when adaptation is necessary and which component should be modified.
\subsection{Reliable LLM-Generated BO Components}
LLM-generated algorithmic components also need to satisfy implementation and optimization requirements before they are deployed. Generated code may contain invalid operations, numerically unstable computations, or structures that are incompatible with the BO engine. In the kernel-discovery experiment, generated kernels are checked before they are evaluated. However, more general BO components may require different forms of verification.
Future research should therefore combine LLM-based generation with automatic validation mechanisms. These mechanisms may include interface checking, numerical testing, constraint verification, and performance prediction before a generated component is deployed.
\subsection{Computational Cost and Adaptation Latency}
Algorithm-level adaptation introduces additional computational overhead beyond the BO loop itself. The adaptation process needs to process task and historical information, generate candidate BO components, and evaluate them before deployment. This process may become costly when many candidate components or previous tasks are considered. This issue is particularly important for wireless DTs because network configurations may change over short time scales. If adaptation is too slow, the discovered BO component may no longer be suitable for the current task when deployed. Future research should therefore investigate lightweight adaptation strategies that balance optimization improvement with adaptation cost and latency.
\section{Conclusions}
This article introduced an LLM-assisted BO framework for algorithm-level self-evolution in wireless DTs. The framework retains a stable BO backbone while allowing selected components to evolve across changing tasks. As a proof of concept, we considered GP kernel discovery for multi-user beam selection. The discovered kernel improves the final BO performance by 19.7\% on the training tasks and 8.7\% on the test tasks compared with the strongest standard kernel baseline. Future work will investigate adaptation under task changes, reliable validation of generated components, and the computational cost and latency of algorithm-level evolution.
\bibliographystyle{IEEEtran}
\bibliography{IEEEexample}
\end{document}